\documentclass[epj,final]{svjour}
\usepackage{latexsym}
\usepackage{amssymb}
\usepackage{graphics}
\usepackage{longtable}
\usepackage{graphicx}
\usepackage{dcolumn}
\usepackage{float}
\usepackage{bm}
\usepackage{mathbbol}
\usepackage{color}

\newcommand{\vf}{v_{\mbox{\tiny F}}}
\begin{document}
\title{Josephson-like currents in graphene for arbitrary time-dependent
potential barriers}
\author{Sergey E. Savel'ev\inst{1} \and Wolfgang H\"ausler\inst{2}\and
Peter H\"anggi\inst{2}
} 
%
%
%
%
%
%
%
\institute{Department of Physics, Loughborough University,
Loughborough LE11 3TU, United Kingdom \and Institut f{\"u}r Physik
Universit{\"a}t Augsburg, D-86135 Augsburg, Germany}
\date{Received: date / Revised version: date}
%
\abstract{
From the exact solution of the Dirac-Weyl equation we find
unusual currents $j_y$ running in $y$-direction parallel to a
time-dependent scalar potential barrier $W(x,t)$ placed upon a
monolayer of graphene, even for vanishing momentum component
$p_y$. In their sine-like dependence on the phase difference of
wave functions, describing left and right moving Dirac fermions,
these currents resemble Josephson currents in superconductors,
including the occurance of Shapiro steps at certain frequencies
of potential oscillations. The Josephson-like currents are
calculated for several specific time-dependent barriers. A
novel type of resonance is discovered when, accounting for the
Fermi velocity, temporal and spatial frequencies match.
%
\PACS{
{05.60.Gg}{Quantum transport} \and
{72.80.Vp}{Electronic transport in graphene} \and
{73.22.Pr}{Electronic structure of graphene} \and
{73.40.Gk}{Tunneling} \and
{78.67.Wj}{Optical properties of graphene}
} 
} 
\maketitle
\section{Introduction}
Growing interest to graphene, see e.g.\ Ref.\
\cite{Novoselov05,review}, is stimulated by many unusual and
sometimes counterintuitive properties of this two dimensional
material. Indeed, graphene supplies charge carriers exhibiting
the pseudo-relativistic dynamics of massless Dirac fermions.
One example of the unusual dynamics of electrons and holes in
graphene is the Klein tunneling phenomenon \cite{paradox} which
occurs with unit probability through arbitrarily high and thick
barriers at perpendicular incidence, irrespective of the
particle energy, in accordance with experiment \cite{paradox-exp}.
In consequence, the question arose of how to control the
electron motion in graphene and hence boosted detailed studies
of Dirac fermions under the influence of various forms of scalar
\cite{superl,super5,left,peeters09b,laser1,laser2,Efetov,pnexperiments}
or vector \cite{magstruc} potentials.

So far, many of works were devoted to studies of graphene
subject to static periodic electric fields, since these
structures known as graphene superlattices \cite{superl,super5}
allow controlling both spectrum and transport properties of
electrons in graphene. For instance, it was shown \cite{super5}
that 1D graphene superlattices have a deep analogy with photonic
crystals formed by alternating right-handed and left-handed
transparent media, similar as the earlier stated analogy
\cite{left} of a p-n junction in graphene to a Veselago lens.
Superlattices of electrostatic periodic potentials can be used
to collimate the directional spread of electron beams in
graphene \cite{peeters09b} so that waves of small
transverse momentum will dominate transport properties.

Applying a time-dependent laser field to a pristine graphene
sample opens an alternative and efficient way
\cite{laser1,laser2,Efetov} to control spectrum and transport
properties of graphene samples. It has been shown that changing
the time dependence of laser fields can mimic \cite{laser2} the
influence of any electrostatic graphene superlattices on the
electron spectrum in graphene. Further, Dirac fermions in
graphene superlattices can aquire an effective mass proportional
to the frequency of an applied laser field, accompanied with an
exponential suppression of chiral tunneling even for
perpendicular incidence upon the barrier \cite{laser2,Efetov},
which is in stark contrast to Klein tunneling occuring in the
absence of the laser field. Studies of how electron transport
in graphene is affected by {\it time-and-space dependent
potentials} are yet limited. Recently, it was shown
\cite{our_prl} that even scalar potential barriers can produce
resonant amplification of reflections when modulated at proper
frequencies. Moreover, an unusual current running parallel to
the barrier $W(x,t)$ in $y$-direction has been predicted
\cite{our_prl} for electrons at zero $y$-component of the
electron momentum.

In this article we study in detail this unusual Josephson-like
current for electrons traveling at zero transverse momentum, $p_y=0$,
accross a time-dependent potential barrier $W(x,t)$, assumed as
homogeneous along the $y$-direction. Explicit calculations
reveal Shapiro steps for properly chosen frequencies and/or
electron momentum in a full analogy to the Josephson current
arising through an irradiated barrier between two
superconductors. We also show that this Josephson-like current
in graphene can assume a non-zero {\sc dc}-component,
resemble the {\sc ac}-Josephson effect, and/or be strongly
enhanced at certain spatio-temporal matching conditions.
Experimental test of our predictions should be within reach of
present day nanostructure design on graphene
\cite{pnexperiments}. Note, that a somewhat related effect, an
unusual ballistic side-jump motion of electrons and holes, has
been predicted \cite{schliemann} to occur in semiconductor
quantum wells as a result of Rashba spin-orbit coupling.

\section{Exact solution for a scalar potential of arbitrary
space and time dependence}
The honeycomb lattice of graphene engenders two copies,
$\tau_z=\pm 1$, of Dirac-Weyl Hamiltonians \cite{kanemele}
\begin{equation}
\label{h0} H_0=\vf\:[\hat\tau_z\hat\sigma_x\hat p_x+\hat\sigma_y\hat p_y]\;,
\end{equation}
centered about two inequivalent Dirac points (``valleys'') $K$
and $K'$ at corners of the hexagonal first Brillouin zone where
electron-hole symmetric bands touch; here Pauli matrices
$\hat\sigma_{x,y,z}$ act on two-component spinors representing
sublattice amplitudes, exhibiting opposite Fermion
helicities, $\:\bm{\sigma}\cdot\bm{p}/p=\pm 1\:$. SU(2)
rotations with respect to the vector $\bm{\hat\tau}$ of three Pauli
matrices allow to continuously transform both copies into one
another \cite{beenakker08} which motivated the terminus
``valleytronics'' for isospin manipulations based on eigenstates
to $\hat\tau_z$, Ref.\ \cite{rycerz07}, in analogy to the
well-known research area of spintronics \cite{wolf01}. Proposals
exist to valley polarize carriers, by means of nanoribbons
terminated by zig-zag edges \cite{rycerz07,akhmerov08,peeters09},
by exploiting trigonal warping at elevated
energies \cite{garcia08}, or by absorbing magnetic
textures \cite{ziegler10}.

Below, we focus on valley polarized situations. Indeed, smooth
electromagnetic or disorder potentials do not couple the two
valleys \cite{ando98}, so that calculations can be done
independently, for either $\tau_z=+1$ or $\tau_z=-1$. Including
now the barrier potential $W(x,t)$ the Dirac equation becomes
\begin{eqnarray}
\label{dir1}
\vf(\tau_z\hat p_x-{\rm i}\hat p_y)\Psi_B+\hbar W(x,t)\Psi_A={\rm
i}\hbar\frac{\partial \Psi_A}{\partial t}\\ \nonumber
\vf(\tau_z\hat p_x+{\rm i}\hat p_y)\Psi_A+\hbar W(x,t)\Psi_B={\rm
i}\hbar\frac{\partial \Psi_B}{\partial t}\;,
\end{eqnarray}
where the wave functions $\Psi_A, \Psi_B$ describe electrons on
either of the hexagonal graphene sublattices, $\vf$ is the
Fermi velocity, and the momentum operator is defined as $(\hat
p_x,\hat p_y)=(-{\rm i}\hbar\partial/\partial
x,-{\rm i}\hbar\partial/\partial y)$. This equation has been solved
analytically for time-independent potentials, for rectangular barriers
\cite{paradox}, for trapezoidal barriers \cite{sonin09}, or for
smooth barriers by the WKB method
\cite{cheianov06,silvestrov07,fistul08}. Additional time
dependent harmonic oscillations have been considered, either of
gate voltages applied to each side of the rectangular barrier
\cite{bjoern07}, or of an electric field imposed parallel to the
barrier \cite{Efetov}, or treated in resonance approximation
\cite{laser2}. Recently, the exact solution for $p_y=0$
\cite{our_prl,solomon} has been derived which allows
\cite{our_prl} to uncover new physical phenomena from
spatio-temporal dynamics.

To keep this article self contained we briefly repeat the
crucial steps to obtain the exact solution of eq.\ (\ref{dir1})
for $p_y=0$ and arbitrary potential $W(x,t)$ acting at positive
times, i.e.\ $\:W(x,t<0)=0\:$. The wave functions $\Psi_A$ and
$\Psi_B$ depend on time $t$ and on the coordinate $x$ across the
barrier, but not on the $y$-coordinate. This simplifies
(\ref{dir1}) to read
\begin{eqnarray}\label{dir2}
-{\rm i}\vf\tau_z\frac{\partial \Psi_B}{\partial x}+ W(x,t)\Psi_A={\rm
i}\frac{\partial \Psi_A}{\partial t}\\ \nonumber
-{\rm i}\vf\tau_z\frac{\partial \Psi_A}{\partial x}+ W(x,t)\Psi_B={\rm
i}\frac{\partial \Psi_B}{\partial t}\;.
\end{eqnarray}

To solve (\ref{dir2}) we use the Ansatz
\begin{equation}\label{ansatz}
\psi_{\pm}(x,t)=\frac{1}{2}\left({{\rm e}^{{\rm i}S_{\pm}(x,t)}\atop
\pm\tau_z{\rm e}^{{\rm i}S_{\pm}(x,t)}}\right)
\end{equation}
where the $\pm$ signs distinguish right and left propagating
solutions. Inserting (\ref{ansatz}) into (\ref{dir2}) results in
\begin{equation}
\partial_tS_{\pm}(x,t)\pm\vf\partial_xS_{\pm}(x,t)+W(x,t)=0\;.
\end{equation}
This first order partial differential equation can be solved by
the method of characteristics \cite{couranthilbert}, yielding
\begin{equation}\label{spm}
S_{\pm}(x,t)=S_{\pm}^{(0)}(x\mp \vf t,0)
-\int_0^t{\rm d}t'\; W(x\mp \vf(t-t'),t')
\end{equation}
explicitly in terms of $W(x,t)$. In view of (\ref{ansatz}) the
term $S_{\pm}^{(0)}(x,0)$ describes the initial wave function
$\psi_{\pm}(x,0)$ at time $t=0$ which, in the absence of the barrier
at $t<0$, can be, for example, a plane wave of wave number $k$
in the $x$-direction, $\:S_{\pm}^{(0)}(x,0)=\pm kx\:$, or some
wave packets. Then eq.\ (\ref{spm}), together with (\ref{ansatz}),
describes the full solution for $t>0$
\begin{eqnarray}\label{timeevolution}
&&\psi(x,t)=a_+(x-\vf t)\left(1\atop \tau_z \right) {\rm e}^{-{\rm
i}\int_0^t{\rm d}t'\; W(x - \vf(t-t'),t')}
\nonumber \\ &&\;\mbox{}+a_-(x+\vf t)\left(1\atop -\tau_z\right) {\rm
e}^{-{\rm i}\int_0^t{\rm d}t'\; W(x + \vf(t-t'),t')}
\end{eqnarray}
where $\:a_{\pm}(x)={\rm e}^{{\rm
i}S_{\pm}^{(0)}(x)}/2=[\Psi_A(x,t=0)\pm\tau_z\Psi_B(x,t=0)]/2\:$
encodes the initial condition.
In particular, if
the wave packet is initially purely right moving, $\:a_-=0\:$,
according to (\ref{timeevolution}), it continues propagating to
the right at times $t>0$ with undistorted density distribution
$\:|a_+(x-\vf t)|^2\:$ without reflection, acquiring at most a
phase factor. However, the situation becomes more intriguing
when we consider a superposition of left and right moving waves.

\section{Current density perpendicular and parallel to the barrier}
Next we evaluate the current density
\begin{eqnarray}
\label{current}
j_x(x,t)=\vf\psi^*(x,t)\tau_z\hat\sigma_x\psi(x,t) \nonumber \\
j_y(x,t)=\vf\psi^*(x,t)\hat\sigma_y\psi(x,t)
\end{eqnarray}
for $p_y=0$. Before coming to remarkably nontrivial consequences
from (\ref{timeevolution}) below, let us first study the
$x$-component of the current density, flowing perpendicular to
the barrier,
\begin{equation}\label{jx}
j_x(x,t)=2\vf\left(|a_+(x-\vf t)|^2-|a_-(x+\vf
t)|^2\right)\;,
\end{equation}
which is solely determined by the initial conditions and is {\it
independent\/} of $W(x,t)$. An initially purely right moving
wave packet, $a_-=0$, generates an undistorted current density
peak $j_x=\vf|a_+(x-\vf t)|^2/2$ moving at $\vf$ towards the
right. According to (\ref{jx}), right and left movers in the
initial wave will just add their contributions to the current
density of opposite sign. This confirms the finding of perfect
Klein tunneling through a barrier $W(x,t)$ of any space and even
any time dependence. Furthermore, the current $j_x$ does not
depend on $\tau_z$, giving the same contribution from states
near both valleys $K$ and $K'$. Regarding the current normal to
the barrier we find so far no unusual effects arising from the
superposition of right and left moving amplitudes.

Surprisingly, although we consider electron momenta $p_y=0$, we
find a nonzero value for the current component $j_y$ parallel to the barrier,
\begin{eqnarray}\label{jy}
j_y(x,t)&=&4\tau_z\vf|a_+(x-\vf t)a_-(x+\vf t)|\nonumber\\
&\times&\sin\left[S_+(x,t)-S_-(x,t)\right]\;.
\end{eqnarray}
While eq.\ (\ref{jy}) vanishes for unidirectional wave packets,
$j_y$ becomes nonzero for superpositions of left {\it and\/}
right moving amplitudes, $\:a_+\ne 0\:$ {\em and\/} $\:a_-\ne
0\:$. Note that any initial density peak arising from some
voltage pulse will generically contain simultaneously left and
right moving amplitudes. It is this superposition which causes
the qualitatively new phenomenon of a current {\it along} the
barrier, exhibiting striking properties as described in the
following. The sine-dependence in (\ref{jy}) is reminiscent of
the Josephson effect where it originates from the spatial
overlap of superconducting order parameters in the leads of a
Josephson junction. Similarly, $j_y$ in graphene originates
here due to overlapping amplitudes of left and right moving
fermions.

Contrary to its $x$-component, the $y$-component of the current
manifests a nontrivial space and time dependence. The latter
also depends on $\tau_z$. Therefore, the best way to observe
$j_y$ is to prepare a valley-polarized system. For non- or
partly polarized situations one should add the current
contribution from the second valley, yielding a total current
$j_y(x,t)={\cal P}j_{y}^{K}(x,t)+(1-{\cal P})j_{y}^{K'}(x,t)$
where $j_{y}^{K}$ and $j_{y}^{K'}$ refer to states near $K$ and
$K'$ points, respectively, and ${\cal P}$ measures the degree of
valley polarization such that ${\cal P}=1$ or $0$ corresponds to
complete polarization and ${\cal P}=1/2$ to the unpolarized
situation. Since the contribution from the other Dirac point
$K'$ can compensate the current from $K$, measuring $j_y$ allows
to determine the degree ${\cal P}$ of valley polarization of a
graphene sample.
\begin{figure}[h]
\resizebox{0.5\textwidth}{!}{\includegraphics{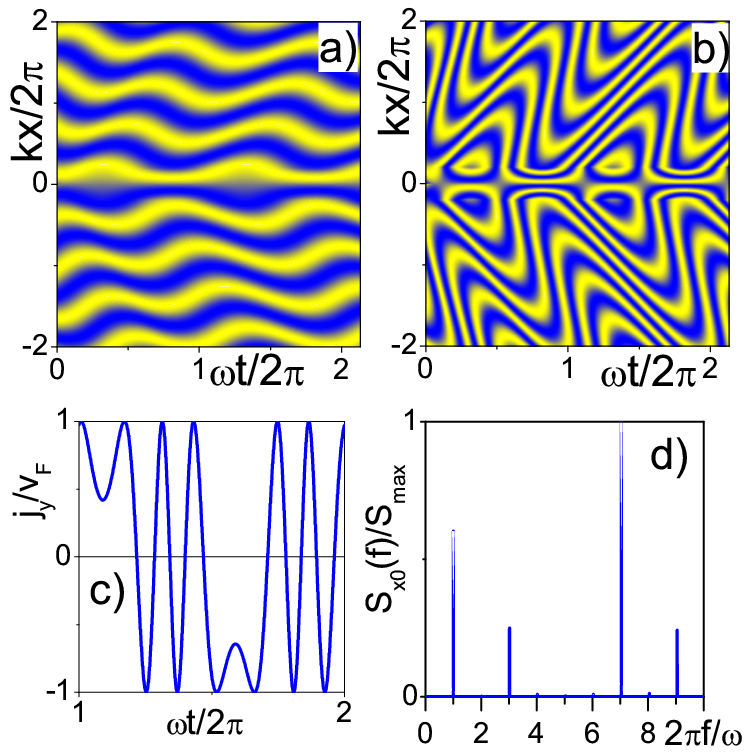}}
\caption{(Color online) Contour plot of $j_y(x,t)$ for the case
of an oscillating square-well potential (\ref{step-pot}).
Bright (yellow): $j_y>0$, dark (blue): $j_y<0$, as calculated
using eq.\ (\ref{jy-step}) for small $W_0/\omega=1.75$, panel
{\sf a)}, and for large $W_0/\omega=10.7$, {\sf b)}. The other
parameters ($\vf/(L\omega)=2.35,\ kL=1.27$) are the same for all
panels {\sf a)--d)}. With growing $W_0$, a wave-like structure
seen in {\sf a)} changes towards a more complicated pattern {\sf
b)} of $j_y$. Panel {\sf c)} display cross sections
$j_y(x_0,t)$ of {\sf b)} at $x_0/L=1$. Current
periodicity seen in {\sf c)} is consistent with the spectrum
${\cal S}_{x_0}(f)$, as defined in (\ref{f-spec}) (normalized
by its maximum value) containing only integer harmonics as shown
in panel {\sf d)}.
}
\label{fig1}
\end{figure}

On the other hand, the current {\it variance\/} in $y$-direction,
\begin{eqnarray}\label{jyvariance}
\langle\hat j_y^2\rangle-\langle\hat j_y\rangle^2&=&\vf^2\Bigl[1
-16|a_+|^2|a_-|^2\\
&\times&\sin^2\left(S_+(x,t)-S_-(x,t)\right)\Bigr]\nonumber\\
&=&\vf^2[1-j_y^2(x,t)/\vf^2]\nonumber
\end{eqnarray}
does not depend on $\tau_z$. In (\ref{jyvariance}) we
have defined the current operator $\:\hat j_y:=\vf\hat\sigma_y\:$. Even
without valley polarization they {\it remain nonzero\/} and {\it
can be measured}.

\section{Josephson-like current flowing along the barrier}\label{results}
For valley-polarized situations, $\tau_z=1$ and ${\cal P}=1$, we
now investigate specific examples $W(x,t)$ and substantiate the
analogy between the current $j_y$ and a superconducting
Josephson current. We consider potentials of amplitudes $W_0$,
varying spatially on lengths scales $L$, and oscillate at
frequency $\omega$ such that they vanish on time average. As
initial condition we assume a superposition of right and left
propagating plane waves of equal amplitudes,
$\:S_{\pm}^{(0)}=\pm kx\:$.

\subsection{Square well of width $2L$}
Let us first consider a single square well barrier
\begin{equation}
W(x,t)=W_0\:\Theta(L-|x|)\:\sin\omega t
\label{step-pot}
\end{equation}
of thickness $2L$ and amplitude $W_0$, oscillating at
frequency $\omega$. Substituting this potential in eq.\
(\ref{spm}) and using eq.\ (\ref{jy}), we derive
\onecolumn
\begin{equation}
\textstyle j_y(x,t)=\vf\:\left\{
\begin{array}{ll}
\textstyle\sin 2\left[kx-\frac{W_0}{\omega}
{\rm sgn}(x)\:\sin^2\left(\frac{\omega t}{2}-
\frac{\omega}{2\vf}\bigl|L-|x|\bigr|\right)\right] &,\quad
t<(L+|x|)/\vf \\
\textstyle\sin 2\left[kx-\frac{W_0}{\omega}{\rm sgn}(x)\:\sin
\left(\frac{\omega}{\vf}\left\{{|x|\atop L}\right\}\right)\:
\sin\left(\omega t-\frac{\omega}{\vf}\left\{{L\atop |x|}\right\}
\right)\right] &,\quad
t\ge(L+|x|)/\vf
\end{array}\right.\;,
\label{jy-step} \end{equation}
\twocolumn
where the curley brackets in the sine arguments of the lower
line refer to $\left\{{|x|<L\atop |x|>L}\right\}$. Fig.\
\ref{fig1} shows $j_y(x,t)$ for small (\ref{fig1}{\sf a}) and
large (\ref{fig1}{\sf b}) amplitude of the potential barrier.
With growing $W_0$ an initial wave-like structure transforms
into a more complicated spatio-temporal pattern. However, as
seen in \ref{fig1}{\sf c} for fixed $x_0$, the current
$j_y(x_0,t)$ remains periodic in time, as confirmed by the
spectrum
\begin{equation}\label{f-spec}
{\cal S}_{x_0}(f)=\left|\int_0^{\infty}{\rm
d}t\;j_y(x_0,t)\exp(2\pi{\rm i} f t)\right|^2
\end{equation}
of $j_y$ at fixed $x_0$, showing peaks in Fig.\ \ref{fig1}{\sf
d} only at integer harmonics.

\subsection{Homogeneous electric field}
\begin{figure}[h]
\resizebox{0.5\textwidth}{!}{\includegraphics{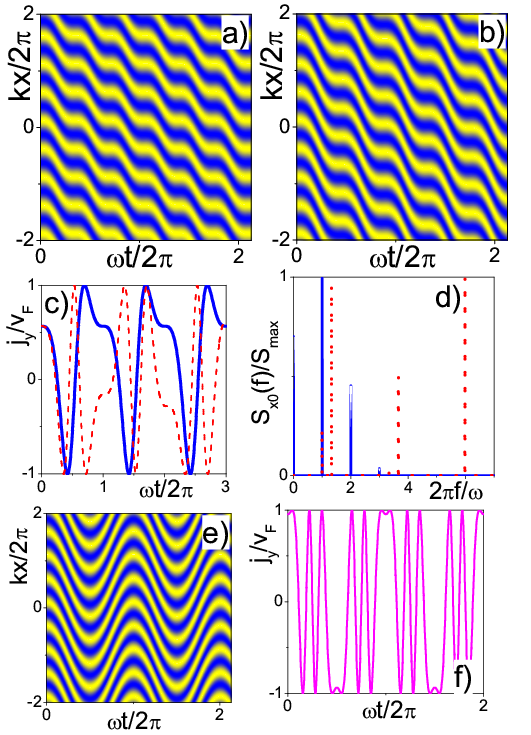}}
\caption{
(Color online) Contour plot of $j_y(x,t)$ for an oscillating
homogeneous electric field (\ref{sin-pot}) at the Shapiro step
$n=2$ corresponding to $W_0/\omega=0.86,\ \vf/L\omega=1.16$,
panel {\sf a)}, and away from Shapiro steps ($W_0/\omega=0.76,\
\vf/L\omega=1.03$), {\sf b)}. Bright (yellow): $j_y>0$, dark
(blue): $j_y<0$, calculated by using eq.~(\ref{jy11}) for
$kL=1.27$. A very regular pattern is seen in {\sf a)} while the
pattern is more ``frustrated'' in {\sf b)}. Panel {\sf c)}
displays cross sections of $j_y(x_0,t)$ at fixed $kx_0=1.27$ for
the first Shapiro step (solid blue line, $W_0/\omega=0.61,\
\vf/L\omega=0.82$) and away from Shapiro steps (dashed red line,
$W_0/\omega=0.75,\ \vf/L\omega=1$). At the Shapiro step, clear
periodic behavior is seen, in contrast to aperiodic oscillations
away from Shapiro steps. This is consistent with the
spectra (normalized w.r.t.\ the peak maximum), cf.\
eq.~(\ref{f-spec}), shown in panel {\sf d)} where only integer
harmonics contribute to the solid blue line while dashed red
contains incommensurate harmonics.
Contour plot of $j_y(x,t)$ for the oscillating homogeneous
electric field (\ref{cos-pot}) {\sf e)}, calculated by using
eq.~(\ref{jy-cos}) for $kL=1.27,\ W_0/\omega=1.75,\
\vf/L\omega=2.35$. Panel {\sf f)} displays the cross section of
{\sf e)} at $x_0=0.5$ which clearly exhibits now always time
periodicity of $j_y(x_0,t)$ for this case.
}
\label{fig2}
\end{figure}
We next consider an {\sc ac}-electric field with amplitude
$W_0/L$ and frequency $\omega$ described by the potential
\begin{equation}
W(x,t)=\frac{W_0x}{L}\sin\omega t\;.
\label{sin-pot}
\end{equation}
For this case, we derive
\begin{equation}\label{jy11}
\textstyle j_y(x,t)=\vf\:\sin 2\left[kx+\frac{W_0\vf}{L}
\left(\frac{t}{\omega}-\frac{\sin\omega
t}{\omega^2}\right)\right]
\end{equation}
from eqs.\ (\ref{spm}) and (\ref{jy}). Now $j_y$ may either
follow the periodicity $\omega$ of (\ref{sin-pot}) or it may
behave aperiodically, compare the 2D contour plots in Fig.\
\ref{fig2}{\sf a} and \ref{fig2}{\sf b}. To see the
non-periodicity of Fig.\ \ref{fig2}{\sf b} more clearly, we plot
in Fig.\ \ref{fig2}{\sf c} cross sections $j_y(x_0,t)$ at fixed
$x_0$ for both cases: the solid blue line, referring to Fig.\
\ref{fig2}{\sf a}, is clearly periodic, while the dashed red
line, referring to \ref{fig2}{\sf b}, is aperiodic.
Corresponding spectra (Fig.\ \ref{fig2}{\sf d}) reveal the
same information: in the solid blue case they contain only
integer harmonics while the non-integer contributions (dashed
red) describe aperiodicity. We can rewrite eq.~(\ref{jy11}) as
a sum
\begin{eqnarray}\label{jy11n}
j_y(x,t)&=&\vf\sum_{n=-\infty}^{\infty}
J_n\left(\frac{2W_0\vf}{\omega^2L}\right)\nonumber\\
&\times&\sin\left(2kx+\frac{2W_0\vf t}{\omega L}-n\omega t\right)
\end{eqnarray}
using Bessel functions $J_n$. The last equation reveals that
at $\omega=\omega_n$ with
\begin{equation}
\omega_n=\sqrt{2W_0\vf/(Ln)}\;,\quad n\in\mathbb{N}
\end{equation}
and integer $n$, the $y$-component of the current exhibits a
peculiarity, similar to the so-called Shapiro steps
\cite{tinkham} of an irradiated Josephson junction. As seen in
Fig.\ \ref{fig2}{\sf c} and \ref{fig2}{\sf d}, frequencies
$\omega=\omega_n$ generate periodic oscillations, which, again
as in the case of Shapiro-steps, can induce a nonzero {\sc
dc}-component of the current at given $x_0$. Here, we remind of
the statement of the previous section, that non-zero {\sc
dc}-currents allow to measure the degree of valley polarization.
In view of eqs.\ (\ref{jy11},\ref{jy11n}) the overall {\sc
dc}-current vanishes after averaging over $x_0$ due to the
harmonic $x$-dependence of $j_y$. Modulating the homogeneous
electric field at $\omega\ne\omega_n$ results in aperiodic
oscillations (Fig.\ \ref{fig2}{\sf b,c,d}) and zero {\sc
dc}-component.

When we consider the same potential, seemingly just phase
shifted in time,
\begin{equation}
W(x,t)=\frac{W_0x}{L}\cos\omega t\:,
\label{cos-pot}
\end{equation}
instead of eq.\ (\ref{jy11}) this yields
\begin{equation}
\textstyle j_y(x,t)=\vf\:\sin 2\left[kx+\frac{2W_0\vf}{\omega^2L}\:
\sin^2\frac{\omega t}{2}\right] \label{jy-cos}
\end{equation}
without a term proportional to $t$ in the square bracket
argument of the sine-functions and therefore without similarity
to Shapiro steps in Josephson junctions. The oscillations of
$j_y$ now are always periodic (as seen in Fig.\ \ref{fig2}{\sf
e}), though $j_y$ could be quite complicated (see for example
Fig.\ \ref{fig2}{\sf f}). For this case there is no {\sc
dc}-component of $j_y$ at any $x$. The reason for this
qualitatively different behaviour as compared to (\ref{jy11})
lies in the discontinuity of (\ref{cos-pot}) at time zero (we
recall that we assume $\:W(x,t<0)=0\:$), contrary to
(\ref{sin-pot}), so that (\ref{cos-pot}) does not simply
correspond to a temporal phase shift. In the limit $\:\omega\to
0\:$ the effect of a homogeneous electric field becomes
time-independent for both forms (\ref{sin-pot}) and
(\ref{cos-pot}).

\subsection{Spatially and temporally periodic potentials}
Now we focus on potentials which are both, periodic in space
(with period $2\pi L$) and in time (with period
$2\pi/\omega$). First consider
\begin{equation}
W(x,t)=W_0\cos(x/L)\:\cos(\omega t)\;.
\label{per-cos-pot}
\end{equation}
Intriguingly, this potential facilitates spatio-temporal mode
matching. In this case, the phase $\phi=S_+-S_-$ describing the
Josephson-like current $j_y\propto \sin(\phi)$ varies as
\begin{eqnarray}\label{non-res}
\phi=S_+-S_-&=&2kx-\frac{4W_0L\vf\sin(\frac{x}{L})}{\omega^2L^2-\vf^{2}}\\
&\times&\sin\left[\frac{\omega L+\vf}{2L}t\right]\:
\sin\left[\frac{\omega L-\vf}{2L}t\right]\;.\nonumber
\end{eqnarray}
In this case, oscillations of $j_y$ persist even when $\omega\rightarrow
0$ since the static potential of spatial periodicity $L$ induces
a frequency component $\vf/L$ to electron waves moving at uniform
velocity $\vf$ which produces phase oscillations
\begin{equation}
\phi=2kx-\frac{4W_0L}{\vf}\sin\left(\frac{x}{L}\right)
\sin^2\left(\frac{\vf}{2L}t\right)\;.
\end{equation}
This reminds of the {\sc ac}-Josephson effect \cite{tinkham}
where {\sc ac}-current oscillations are generated by a
time-independent voltage. By contrast, the potential
\begin{equation}\label{per-sin-pot}
W(x,t)=W_0\cos(x/L)\:\sin(\omega t)
\end{equation}
is continuous in time, yielding a phase
\begin{equation}
\phi=2kx-\frac{2W_0L\sin(\frac{x}{L})}{\omega^2L^2-\vf^2}
\left(\omega L\sin\left(\frac{\vf t}{L}\right)-\vf\sin(\omega t)\right)
\end{equation}
that vanishes when $\:\omega\to 0\:$ with no {\sc
ac}-Josephson-like effect. The essential difference (including
presence or absence of the {\sc ac}-Josephson-like effect) in
the time dependence of $j_y$ for (\ref{per-cos-pot}) and
(\ref{per-sin-pot}) is again related to the continuity of the
potentials at $t=0$.
\begin{figure}[h]
\resizebox{0.5\textwidth}{!}{\includegraphics{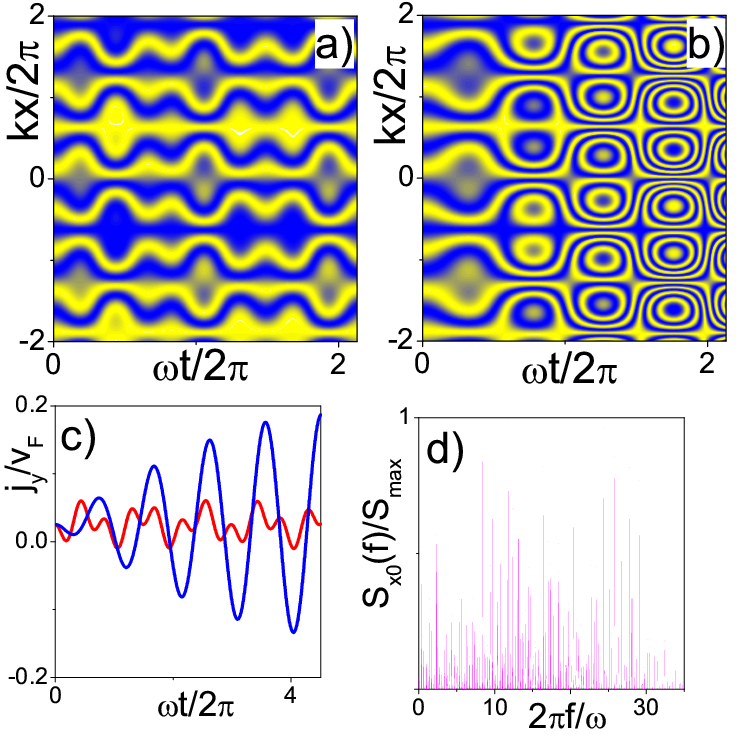}}
\caption{
(Color online) Contour plot of $j_y(x,t)$ for an oscillating
periodic potential (\ref{per-cos-pot}), away from
spatio-temporal matching resonance {\sf a)}, cf.\
eq.~(\ref{non-res}) ($\vf/L\omega=2.35,\ W_0/\omega=1.75$) and
at resonance {\sf b)}, cf.\ eq.~(\ref{node2}) ($\vf/L\omega=1,\
W_0/\omega=1.75$). Bright (yellow): $j_y>0$, dark (blue):
$j_y<0$. Near resonance, amplitude and frequency of
$j_y(x_0,t)$-oscillations increase with time, as seen in {\sf
c)} for fixed $kx_0=0.0127$ for the nearly resonant case in blue
($\vf/L\omega=1.1,\ W_0/\omega=1.75$), compared to the case away
from resonance in red ($\vf/L\omega=2.35,\ W_0/\omega=1.75$),
where oscillations stay small. Panel {\sf d)} shows the
spectrum for large amplitude $W_0/\omega=10.7$ of the potential
(\ref{per-cos-pot}) at $kx_0=1.27$ for the non-resonant
situation $\vf/L\omega=2.35$: now many harmonics contribute,
also at frequencies much higher than $\omega/2\pi$. Such types
of spectra are known \cite{rakh} to enable signal amplification.
} 
\label{fig3}
\end{figure}

On the other hand, when $\:\omega\rightarrow\vf/L$,
spatio-temporal matching occurs so that both of the previous
solutions vary proportional to $t$, resulting in Josephson-like
currents
\begin{equation}\label{node1}
j_y=\vf\sin\Bigl(2kx- t\,W_0\sin(x/L)\sin(\omega t)\Bigr)
\end{equation}
for potential (\ref{per-cos-pot}) and
\begin{equation}\label{node2}
j_y=\vf\sin\Bigl(2kx+ t\,W_0\sin(x/L)\cos(\omega t)\Bigr)
\end{equation}
for the potential in (\ref{per-sin-pot}). As a result, $j_y$
amplifies with time. Indeed, for small values $W_0|\sin(x_0/L)|$ the
amplitude of $j_y(x_0,t)$ oscillations grows resonantly within
times $t\lesssim 2\pi/(W_0|\sin(x_0/L)|)$, before it saturates,
while the ``effective'' frequency of the oscillations keeps
increasing with time. We mention here the analogy to resonant
excitations of plasmonic oscillations (Wood's anomaly
\cite{woods}) by spatio-temporal matching of the incident light
with the grating period. Fig.\ \ref{fig3} depicts contour
plots of $j_y(x,t)$ to illustrate this spatio-temporal mode
matching for potential (\ref{per-cos-pot}): panel \ref{fig3}{\sf
a} away from resonance and \ref{fig3}{\sf b} at resonance,
$\omega=\vf/L$. In the latter case the initially slowly varying
structure is seen to ``accelerate'' as time increases. Resonant
amplification of $j_y(x_0,t)$ is shown in Fig.\ \ref{fig3}{\sf
c} for fixed $x_0$ and small $W_0|\sin(x/L)|$ (blue line); away
from resonance (red line) $j_y(x_0,t)$ stays small.

While at small $W_0$ only few harmonics contribute to the
spectrum of $j_y(x_0,t)$ their number and also the corresponding
frequency range considerably increases at large $W_0$,
particularly in the non-resonant case $\omega\ne\vf/L$. This is
demonstrated in Fig.\ \ref{fig3}{\sf d}. Those types of
spectra, containing dense frequency components over a wide range
of frequencies, can be employed for parametric amplification of
a weak signal (encoded in small variations of the amplitude
$W_0$) by a strong drive (large amplitude $W_0$) \cite{rakh}.

Finally, we discuss a {\it spatial\/} Shapiro-step
peculiarity arising in (\ref{non-res}) due
to the interplay of a linearly increasing term $\sim kx$ and an
oscillatory term $\sim\sin(x/L)$ in $\phi$. At fixed time $t_0$
the current density $j_y(x,t_0)\propto\sin\phi$ becomes
spatially periodic whenever the Shapiro-step condition
\begin{equation}\label{spatial-shapiro}
k=k_n=\frac{n}{L}\quad,\quad n\in\mathbb{N}
\end{equation}
is met for the $x$-component of the electron momentum $k$.
Otherwise, $j_y(x,t_0)$ behaves aperiodic in space.
Interestingly, the resonance condition (\ref{spatial-shapiro})
can imply a nonzero spatial average $\langle j_y\rangle(t_0)$
for the current density of electrons with momentum $k=k_n$.

\subsection{Traveling wave potential}
\begin{figure}[h]
\resizebox{0.5\textwidth}{!}{\includegraphics{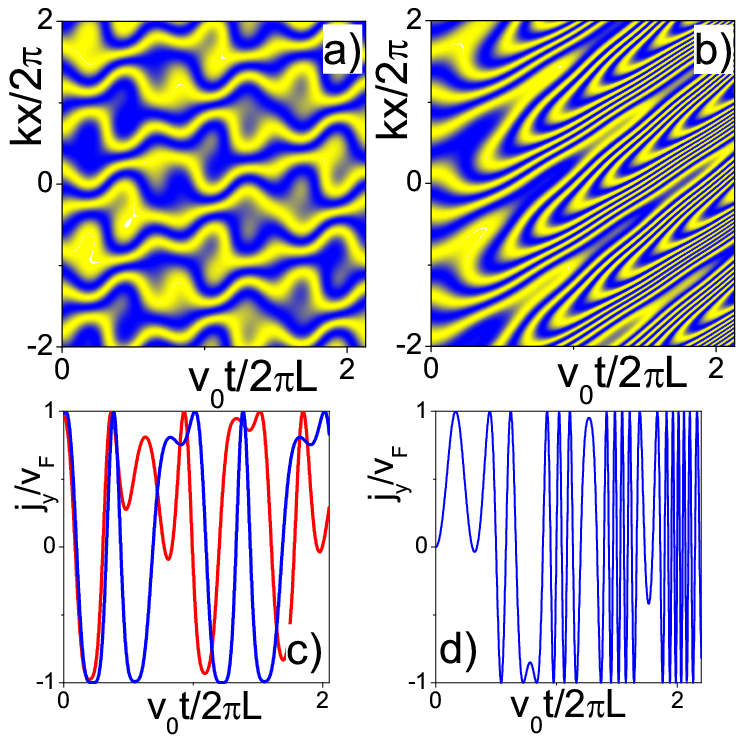}}
\caption{
(Color online) Contour plot of $j_y(x,t)$ for the traveling wave
potential (\ref{travel-wave}), away from the resonance of
matching velocities ($\vf/v_0=2.35$) {\sf a)}, cf.\
eq.~(\ref{run0}) and at the resonance ($\vf/v_0=1$) {\sf b)},
cf.\ eq.~(\ref{run1}). Bright (yellow): $j_y>0$, dark (blue):
$j_y<0$. Throughout Fig.\ \ref{fig4} we use parameters
$kL=1.27$ and $W_0L/v_0=1.75$. Figures {\sf c)} and {\sf d)}
display the time dependence of $j_y(x_0,t)$ at fixed $x_0=L$
according to eq.~(\ref{run0}): periodic oscillations (blue) are
seen in the commensurate case, $\vf/v_0=2$, and aperiodic
oscillations (red) when $\vf/v_0= 2.35$. The resonant case
$\vf/v_0=1$ is shown in {\sf d)}: according to eq.~(\ref{run1})
the effective frequency of the oscillations now increases with
time.
}
\label{fig4}
\end{figure}

Let us finally consider a potential
\begin{equation}\label{travel-wave}
W(x,t)=W_0\sin\frac{x-v_0t}{L}\;,
\end{equation}
describing a traveling wave which can be generated by running
monochromatic electromagnetic waves. Using equation (\ref{spm})
we derive
\begin{eqnarray}\label{run0}
&&\phi=2\Biggl[kx-\frac{W_0L}{(v_0^2-\vf^2)}
\Bigl[\vf\Bigl(\cos\frac{x-v_0t}{L}\\
&&\mbox{}-\cos\frac{x-\vf
t}{L}\Bigr)-(v_0-\vf)\sin \frac{x}{L}\:\sin
\frac{\vf t}{L}\Bigr]\Biggr]\;,\nonumber
\end{eqnarray}
from which we expect a competition between the velocities $v_0$
and $\vf$. In the resonant case, $\:v_0\rightarrow\vf\:$, the
phase $\phi=\:S_+-S_-\:$ grows proportional to time $t$,
\begin{equation}\label{run1}
\phi=2kx-t\;W_0\sin\frac{x-\vf t}{L}+\frac{W_0L}{\vf}
\sin\frac{x}{L}\sin\frac{\vf t}{L}
\end{equation}
and we find again a behavior resembling the Wood's anomaly.
However, from comparing equations (\ref{node1}) and
(\ref{node2}) with equation (\ref{run1}) is seen that the
non-propagating wave (\ref{per-cos-pot}) comes with node-like
points in space, where $\sin(x/L)=0$ and the dynamics of
Josephson-like current is frozen. By contrast, the running wave
potential (\ref{travel-wave}) shows non-trivial dynamics of
$j_y$ everywhere (there are no nodes of $j_y$ in this case). As
a result, the contour plots Fig.\ \ref{fig4}{\sf a,b} resemble
the corresponding distributions Fig.\ \ref{fig3}{\sf a,b} when
tilted by 45$^\circ$ degrees. When $k=k_n$
(\ref{spatial-shapiro}), $j_y(x,t_0)$ becomes periodic in space
at fixed time $t_0$. As for the non-propagating potentials
(\ref{per-cos-pot},\ref{per-sin-pot}) we find spatial
Shapiro-steps in this case for any $v_0$ and $\vf$.
When $v_0$ and $\vf$ are commensurate (but not equal),
$j_y(x_0,t)$ becomes a periodic function in time at given $x_0$.
This is seen in Fig.\ \ref{fig4}{\sf c}. For $v_0=\vf$ the
already mentioned resonant case arises where the effective
frequency of $j_y(x_0,t)$ increases with time, cf.\
(\ref{run1}), as depicted in Fig.\ \ref{fig4}{\sf d}.

\section{Conclusions}
Using the exact solution of the Dirac equation for electrons in
graphene moving perpendicular to a scalar potential barrier
$W(x,t)$, we calculate the current component $j_y$ parallel to
the barrier. In valley polarized situations for packets
containing both, left and right moving waves, this current is
nonzero despite of the vanishing incident momentum $p_y$. Its
variance remains nonzero even without valley polarization. The
here predicted current in graphene strikingly resembles the
Josephson current of coupled superconductors and we find
solutions that resemble Shapiro steps. Both, temporal and
spatial Shapiro steps have been established, exhibiting nonzero
mean current when averaged w.r.t.\ time or space. For time
oscillating graphene superlattices and for traveling wave
potentials, resonances were predicted due to spatio-temporal
matching which can strongly amplify Josephson-like currents in
graphene and, at large driving amplitudes, can generate a broad
range of dense frequency components in the spectrum of
$j_y(x_0,t)$ at a given $x_0$. A resemblence to the {\sc
ac}-Josephson effect can arise.

\begin{acknowledgement}
SES acknowledges support from the Alexander von Humboldt foundation
through the Bessel prize and thanks Sasha Alexandrov and Viktor
Kabanov for stimulating discussions. Also, SES was partially 
supported by Ministry of Science of Montenegro, under Contract No 01-682.
PH thanks the Nano-Initiative Munich (NIM).
\end{acknowledgement}


\begin{thebibliography}{}
\bibitem{Novoselov05} K.S. Novoselov, A.K. Geim, S.V. Morozov, D. Jiang,
M.I. Katsnelson, I.V. Grigorieva, S.V. Dubonos, A.A. Firsov,
Nature {\bf 438}, (2005) 197.
\bibitem{review} A.H. Castro Neto, F. Guinea, N.M.R. Peres, K.S. Novoselov,
A.K. Geim, Rev.\ Mod.\ Phys.\ {\bf 81}, (2009) 109.
\bibitem{paradox} M.I. Katsnelson, K.S. Novoselov, A.K. Geim, Nature Phys.\
{\bf 2}, (2006) 620.
\bibitem{paradox-exp} N. Stander, B. Huard, D. Goldhaber-Gordon,
Phys.\ Rev.\ Lett.\ {\bf 102}, (2009) 026807;
A.F. Young, P. Kim, Nature Phys.\ {\bf 5}, (2009) 222;
S.-G. Nam, D.-K. Ki, J.W. Park, Y. Kim, J.S. Kim, H.-J. Lee,
Nanotechnology {\bf 22}, (2011) 415203.
\bibitem{superl} C.X. Bai, X.D. Zhang, Phys.\ Rev.\ B {\bf 76}, (2007) 075430;
C.H. Park, L. Yang, Y.W. Son, M.L. Cohen, S.G. Louie,
Nature Physics {\bf 4}, (2008) 213;
C.H. Park, L. Yang, Y.W. Son, M.L. Cohen, S.G. Louie,
Phys.\ Rev.\ Lett.\ {\bf 101}, (2008) 126804;
M. Barbier, P. Vasilopoulos, F.M. Peeters, Phys.\ Rev.\ B {\bf 81}, (2010) 075438;
L.Z. Tan, C.H. Park, S.G. Louie, Phys.\ Rev.\ B {\bf 81}, (2010) 195426.
\bibitem{super5} Y.P. Bliokh, V. Freilikher, S. Savel'ev, F. Nori,
Phys.\ Rev.\ B {\bf 79}, (2009) 075123.
\bibitem{left} V.V. Cheianov, V. Fal'ko, B.L. Altshuler, Science {\bf 315}, (2007) 1252;
V.A. Yampol'skii, S. Savel'ev, F. Nori, New J.\ Phys.\ {\bf 10}, (2008) 053024.
\bibitem{peeters09b} M. Barbier, P. Vasilopoulos, F.M. Peeters,
Phys.\ Rev.\ B {\bf 80}, (2009) 205415.
\bibitem{laser1} H.L. Calvo, H.M. Pastawski, S. Roche, L.E.F. Foa Torres,
Appl.\ Phys.\ Lett.\ {\bf 98}, (2011) 232103.
\bibitem{laser2} S.E. Savel'ev, A.S. Alexandrov, Phys.\ Rev.\ B {\bf 84}, (2011) 035428.
\bibitem{Efetov} M.V. Fistul, K.B. Efetov, Phys.\ Rev.\ Lett.\ {\bf 98}, (2007) 256803.
\bibitem{pnexperiments} H.-Y. Chiu, V. Perebeinos, Y.-M. Lin, P. Avouris,
Nano Lett.\ {\bf 10}, (2010) 4634;
M.Y. Han, B. \"Ozyilmaz, Y. Zhang, P. Kim,
Phys.\ Rev.\ Lett.\ {\bf 98}, (2007) 206805;
B. Huard, J.A. Sulpizio, N. Stander, K. Todd, B. Yang, D. Goldhaber-Gordon,
Phys.\ Rev.\ Lett.\ {\bf 98}, (2007) 236803;
B. \"Ozyilmaz, P. Jarillo-Herrero, D. Efetov, D.A. Abanin, L.S. Levitov, P. Kim,
Phys.\ Rev.\ Lett.\ {\bf 99}, (2007) 166804;
J.R. Williams, L. DiCarlo, C.M. Marcus, Science {\bf 317}, (2007) 638.
\bibitem{magstruc} T.K. Ghosh, A. De Martino, W. H\"ausler, L. Dell'Anna,
R. Egger, Phys.\ Rev.\ B {\bf 77}, (2008) 081404(R);
W. H\"ausler, A. De Martino, T.K. Ghosh, R. Egger, Phys.\ Rev.\ B {\bf 78}, (2008) 165402;
W. H\"ausler, R. Egger, Phys.\ Rev.\ B {\bf 80}, (2009) 161402(R);
E. Grichuk, E. Manykin, Eur.\ Phys.\ J.\ B {\bf 86}, (2013) 210.
\bibitem{our_prl} S.E. Savel'ev, W. H\"ausler, P. H\"anggi, Phys.\ Rev.\ Lett.\
{\bf 109}, (2012) 226602.
\bibitem{schliemann} J. Schliemann, Phys.\ Rev.\ B {\bf 75}, (2007) 045304.
\bibitem{kanemele} C.L. Kane, E.J. Mele, Phys.\ Rev.\ Lett.\ {\bf 95}, (2005) 226801.
\bibitem{beenakker08} C.W.J. Beenakker, Rev.\ Mod.\ Phys.\ {\bf 80}, (2008) 1337.
\bibitem{rycerz07} A. Rycerz, J. Tworzydlo, C.W.J. Beenakker,
Nature Physics {\bf 3}, (2007) 172.
\bibitem{wolf01} S.A. Wolf, D.D. Awschalom, R.A. Buhrman, J.M. Daughton,
S. von Moln\'ar, M.L. Roukes, A.Y. Chtchelkanova, D.M. Treger,
Science {\bf 294}, (2001) 1488.
\bibitem{akhmerov08} A.R. Akhmerov, J.H. Bardarson, A. Rycerz,
C.W.J. Beenakker, Phys.\ Rev.\ B {\bf 77}, (2008) 205416.
\bibitem{peeters09} J.M. Pereira, F.M. Peeters, R.N. Costa Filho,
G.A. Farias, J.\ Phys.: Condens.\ Matter {\bf 21}, (2009) 045301.
\bibitem{garcia08} J.L. Garcia-Pomar, A. Cortijo, M. Nieto-Vesperinas,
Phys.\ Rev.\ Lett.\ {\bf 100}, (2008) 236801.
\bibitem{ziegler10} A. Hill, A. Sinner, K. Ziegler, New J.\ Phys.\ {\bf 13}, (2011) 035023.
\bibitem{ando98} T. Ando, T. Nakanishi, R. Saito,
J.\ Phys.\ Soc.\ Jpn.\ {\bf 67}, (1998) 2857.
\bibitem{sonin09} E.B. Sonin, Phys.\ Rev.\ B {\bf 79}, (2009) 195438.
\bibitem{cheianov06} V.V. Cheianov, V.I. Fal'ko, Phys.\ Rev.\ B {\bf 74}, (2006) 041403(R).
\bibitem{silvestrov07} P.G. Silvestrov, K.B. Efetov,
Phys.\ Rev.\ Lett.\ {\bf 98}, (2007) 016802.
\bibitem{fistul08} S.V. Syzranov, M.V. Fistul, K.B. Efetov,
Phys.\ Rev.\ B {\bf 78}, (2008) 045407.
\bibitem{bjoern07} B. Trauzettel, Ya.M. Blanter, A.F. Morpurgo,
Phys.\ Rev.\ B {\bf 75}, (2007) 035305.
\bibitem{solomon} D. Solomon, Can.\ J.\ Phys.\ {\bf 88}, (2010) 137.
\bibitem{couranthilbert} R. Courant and D. Hilbert, {\it Methods of Mathematical Physics\/}
(Wiley-Interscience, 1962) Volume II.
\bibitem{tinkham} M. Tinkham, {\it Introduction to Superconductivity\/}
(Dover Publications Inc., 2004).
\bibitem{rakh} S. Savel'ev, A.L. Rakhmanov, F. Nori,
Phys.\ Rev.\ E {\bf 72}, (2005) 056136; S. Savel'ev, A.M. Zagoskin,
A.L. Rakhmanov, A.N. Omelyanchouk, Z. Washington, F. Nori,
Phys.\ Rev.\ A {\bf 85}, (2012) 013811.
\bibitem{woods} H. Raether, {\it Surface Plasmons\/} (Springer, New York, 1988).
\end{thebibliography}
\end{document}